\documentclass[baaa]{baaa}

\usepackage[pdftex]{hyperref}
\usepackage{subfigure}
\usepackage{natbib}
\usepackage{helvet,soul}
\usepackage[font=small]{caption}

\contriblanguage{1}

\contribtype{2}

\thematicarea{2}

\title{Fractal Gravitation }

\titlerunning{Fractal Gravitation}

\author{T. Canavesi\inst{1,2}}

\authorrunning{Canavesi.}

\contact{tobiascanavesi@gmail.com}

\institute{
Instituto de Física de La Plata, CONICET{--}UNLP{, Argentina} 
\and 
Facultad de Ciencias Astron\'omicas y Geof\'isicas{, UNLP, Argentina}
}

\resumen{
Considerando los datos de GAIA para {$\approx 10^6$} estrellas alrededor del {baricentro,} estimamos la dimensión fractal para diferentes regiones en la Via Láctea.	Luego utilizamos esas dimensiones fractales para calcular el potencial gravitacional considerando al medio como un fractal continuo. Por {último,} utilizamos el potencial gravitacional para derivar la velocidad circular {y ajustar} la curva de rotación en la Vía Láctea. Para ello, utilizamos dos modelos numéricos, el primero considerando densidad uniforme y el segundo, más realista, de núcleo y disco. En ninguno de los modelos {consideramos} materia oscura. Estudiamos su validez contrastándolos con los datos de velocidad circular de la Vía Láctea.
}

\abstract{
Considering the GAIA data for {$\approx 10^6$} stars around the {barycenter,} we estimate the fractal dimension for different regions in the Milky Way. Then we use those fractal dimensions to calculate the gravitational potential considering the medium as a continuous fractal. Finally, {we use the gravitational potential to infer} the circular velocity {and adjust} rotation curves in the Milky Way. {For this,} we use two numerical models, the first considering uniform density and a second more realistic of a bulge and a disk. In none of these models we consider dark matter. We study their validity comparing them with circular speed data from the Milky Way.
}


\begin{document}

\maketitle

\section{Introduction}\label{S_intro}

In the first part of the twentieth century mathematics was concerned with sets that are sufficiently regular, and functions over them, to which classical calculus can be applied. But in many relevant situations, irregular objects provide a better representation of natural phenomena.  In such cases Fractal geometry is a better tool to deal with the real world irregularities than Euclidean geometry.

{It} is known that star formation regions in galaxies have a fractional dimension \citep{2000ApJ...530..277E,2001AJ....121.1507E} with $D \approx 2.3$. In others words the regions occupied by matter can be considered as a fractal embedded in 3 dimensional Euclidean geometry.

{In} this work we consider the matter distribution in the Galaxy as a fractal media, and use fractional integrals to calculate the Newtonian potential, following the {work by \citet{Muslih}}. We use two numerical models, the first one considering uniform density and a second more realistic one of a bulge and a disk. In none of these models we consider dark matter. Finally we contrasts our results again the rotation curve of the Milky Way.

\section{Data}

Gaia is a mission of the European Space Agency (ESA) that provides radial velocity and position measurements for more than one billion stars in our Galaxy and the entire Local Group. 

The data of the Gaia mission (Data Release 2) was used to obtain the Cartesian coordinates of {$\approx  10^6$} stars \citep{2018AJ....156...58B}.

\section{Box counting and mass dimension of fractal systems}

A fractal is a set for which the Hausdorff-Besicovitch dimension is not equal to the topological dimension. The Hausdorff-Besicovitch dimension is not practical to compute, so alternative definitions are used. The most common is the box-counting dimension. For a subset of points $F \in \mathbf{R}^n$ the definition is

\begin{equation}\label{boxd}
dim_B F = \lim_{\delta\to 0} \frac{\log N_\delta (F)}{-\log \delta}\ ,
\end{equation}

\noindent
{where $N_\delta(F)$ is} the number of $\delta$-mesh cubes that intersect $F$. 

{{Eq.~}\eqref{boxd}} requires the size of the mesh $\delta$ to vanish. However, in real systems the fractal structure of the media cannot be observed at all scales. In general, physical systems have a minimal length scale $R_{0}$, which is the smallest size from which we can regard the structure as a fractal. In our case $R_{0}=147~\text{pc}\sim4.53\times10^{18}~\text{m}$. 

{We therefore} need a physical analog to Eq.~\eqref{boxd}. For this we introduce the mass dimension, based on the idea of how the mass of a system scales with the system size, considering unchanged density \citep{TarasovBook}.

{Let} $M(W)$ be the mass of a region $W$ of the medium of characteristic size R. The mass dimension is defines as

\begin{equation}
\label{massD}
M_{D}\sim \left(\frac{R}{R_{0}}\right)^{D}\,.
\end{equation}

{By} taking the logarithm of this formula we can show that $D$ approximates $dim_B F$ as long as $R\gg R_{0}$. From now on we use the terms ``box counting dimension'' and ``mass dimension'' interchangeably.

{The} mass dimension characterizes how the system fills the Euclidean space. If we assume that matter is distributed over a fractal with constant density, then the mass enclosed in a volume of characteristic size $R$ satisfies the power-law Eq.~\eqref{massD}, with non integer $D$, whereas for a regular $n$-dimensional Euclidean object $D=n$. So a fractal medium is a medium with non integer mass dimension.

\subsection{Box counting dimension for the milky way}\label{dimensionmilky}

We developed a R code to calculated this dimension in 2D and 3D, avoiding both boundary and small data set problems.

\begin{figure}[t]
	\centering
	\includegraphics[width=0.4\textwidth]{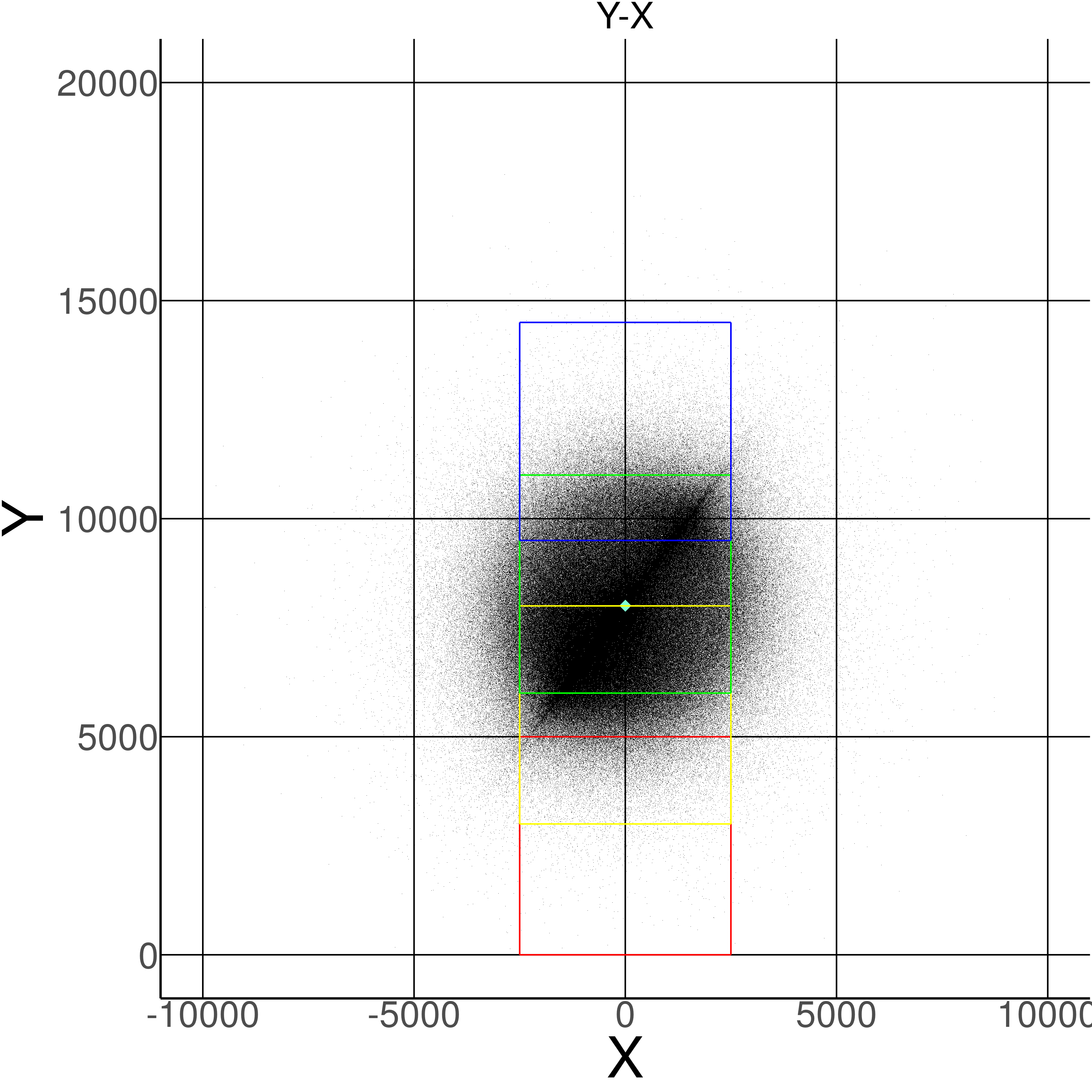}
	\caption{A slice in the X-Y plane from the set of data. {In color, different} cubes are plotted with a difference of 500~pc to determine local variation of the box counting {dimension. The red cube starts} in the center of the {Milky Way, while the blue one  ends at about 13.5~kpc}.}
	\label{cubosslice}
\end{figure}

{We used} our code to calculate the box counting dimension of a cube of {$5~\text{kpc}$ sides, including the $\approx  10^6$ stars of the work by \citet{2018AJ....156...58B}. We consider different} boxes with a step of $100~\text{pc}$ outward from the {Milky Way center,} where each box slides over the 3D data overlapping the previous {placement. Fig.~\ref{cubosslice}} illustrates the pattern of scanning. We find that fractal dimension is changing from $D=2.3$ to $D=2.7$ being very close to $D=3$ when the center of the cube is about $5~\text{kpc}$ from the galactic center (see Fig.~\ref{dimension}). To {establish} this we consider the {galactocentric reference} frame, which requires specifying the {distance from the Sun to the galactic center. The} default position of the {galactic center} in {international celestial reference system (ICRS)} coordinates was {taken from \citet{Reid_2004}}, and the distance to the galactic center is {set to $8.3~\text{kpc}$} \citep{Gillessen_2009}.

\section{Fractal dimension and rotation curves}

\subsection{Mass distribution on fractals}

The mass on a set $W\subset\mathbb{R}^3$ distributed with density $\bar{\rho}(\bar{\textbf{r}},t)$ is defined by

\begin{equation}\label{eq:3mass}
M_{3}(W)=\int_{W}\bar{\rho}(\bar{\textbf{r}},t){\,\text{d}}\bar{V_{3}},
\end{equation}

\noindent
{where $\text{d}\bar{V_{3}}=\text{d}\bar{x}\,\text{d}\,\text{d}\bar{y}\,\text{d}\bar{z}$} for Cartesian coordinates $\bar{x}$,$\bar{y}$,$\bar{z}$. Introducing the dimensionless variables $x=\bar{x}/R_{0}$, $y=\bar{y}/R_{0}$, $z=\bar{z}/R_{0}$, $\textbf{r}=\bar{\textbf{r}}/R_{0}$, where $R_{0}$ is the aforementioned characteristic scale, and the density $\rho(\textbf{r},t)={R_{0}^3\,\bar{\rho}}(\textbf{r}R_{0},t)$ with units of mass, we obtain

\begin{equation}\label{eq:3mass2}
M_{3}(W)=\int_{W}\rho(\textbf{r},t){\,\text{d}}V_{3},
\end{equation}

\noindent
{where $\text{d}V_{3}=\text{d}x\,\text{d}y\,\text{d}z$}. This representation allows us to generalize Eq.~\eqref{eq:3mass2} to fractal media and fractal distribution of mass, as follows. {Let us} consider a mass distribution on a metric set $W$ with fractional dimension $D$, with density function $\rho(\textbf{r},t)$, then the mass is defined as \citep{TarasovBook}

\begin{equation}\label{eq:Dmass}
M_{D}(W)=\int_{W}\rho(\textbf{r},t){\,\text{d}}V_{D},
\end{equation}

\noindent
{where} $\textbf{r}$, $x$, $y$ and $z$ are dimensionless variables, so $\rho(\textbf{r},t)$ has units of mass, and

\begin{equation}\label{eq:dV_D}
dV_{D}=c_{3}(D,\textbf{r}){\,\text{d}}V_{3},
\end{equation}

\begin{equation}\label{eq:c_3}
c_{3}(D,\textbf{r})\propto |x|^{\alpha_{1}-1}|y|^{\alpha_{2}-1}|z|^{\alpha_{3}-1}\ ,
\end{equation}

\noindent
{with} $D=\alpha_{1}+\alpha_{2}+\alpha_{3}$. {Here,} $c_{3}(D,\textbf{r})$ is the density of the points of $W$ in the Euclidean space $\mathbb{R}^3$, the form of which is defined by the symmetries of the fractal medium. The overall numerical factor will not affect the final {results.}

{For} $\rho(\textbf{r})=\rho(|\textbf{r}|)$, we have $\alpha_{1}=\alpha_{2}=\alpha_{3}=D/3$, implying for a homogeneous medium $\rho(\textbf{r})=\rho_{0}=const$ on a ball $W=\{ \textbf{r}:|\textbf{r}|\leq R \}$

\begin{eqnarray}
M_{D}(W)=\alpha\, \rho_{0}\int_{W}|\textbf{r}|^{D-1}{\,\text{d}}|\textbf{r}|\,,
\end{eqnarray}

\noindent
{where} $\alpha$ is a proportionality factor. As a result, we have $M(W)\sim R^{D}$, {\it i.e.} we derive Eq.~\eqref{massD} up to the numerical factor. This allows us to describe the fractal medium with non-integer mass dimension $D$. {Eq.~}\eqref{eq:Dmass} was used to describe fractal media in the framework of fractional continuous model \citep{TARASOV2005a,TARASOV2005b}.

\subsection{Fractal potential}
The central proposal of the present note is to replace the solution of Poisson equation in three dimensions by the corresponding solution on a fractal set, given {by \citet{Muslih}}

\begin{equation}
\label{phifractal}
\phi(\textbf{r})=- G\gamma\int_{W}\frac{\rho(\textbf{r}')}{|\textbf{r}-\textbf{r}'|^{D-2}}{\,\text{d}V'_{D}},
\end{equation}

\noindent
{where} $\textbf{r}$ and $\textbf{r}'$ are dimensionless radius vectors and $D$ the fractional mass dimension of the matter distribution. The dimensionful proportionality constant $\gamma$ is given by
\begin{equation}\label{gamma}
	\gamma=\frac{\Gamma[(D-2)/2]}{\pi^\frac{D-2}{2} R_0^{D-2}}.
\end{equation}

{We assume that gravity propagates on the fractal defined by the matter distribution. This is similar to a gravity localization effect \citep{randallsundrum}}. Notice that in $D=3$ we recover the standard Newtonian form of the potential. 

\subsection{Rotation curves}

The rotation curves of spiral galaxies are one of the best tools to determine their mass distribution, they also provide fundamental information to understand their dynamics. To {find} out how our model fits the data, we use the circular velocity given by $v^2_c(r)=r{\,\frac{\text{d}\phi}{\text{d}r}}$.

\section{The fit in {the Milky Way}}\label{milkyway_fit}

\begin{figure}
	\centering
	\includegraphics[width=.42\textwidth]{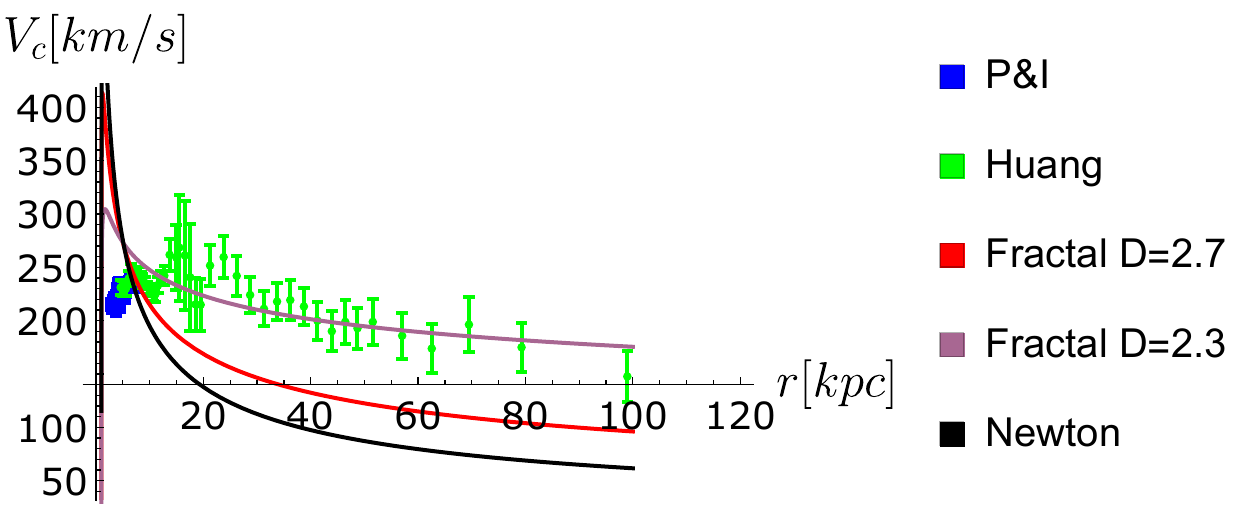}
	\caption{{Results of the uniform density bulge model. Velocity data of \citet{2016MNRAS.463.2623H} (green squares) and \citet{2017SoftX...6...54P} (blue squares). The color curves indicate the fits using the fractal dimension {$D=2.7$ (red) and $D=2.3$ (purple), as well as the} Newtonian case $D=3$ (black). In all cases the mass of the bulge is set to $M_b=6.29 \times 10^{10}~\text{M}_\odot$ and its size to} $R_{b}=1.08~\text{kpc}$.}
	 \label{milky}
\end{figure}

\begin{figure}
	\centering
	\includegraphics[width=0.44\textwidth]{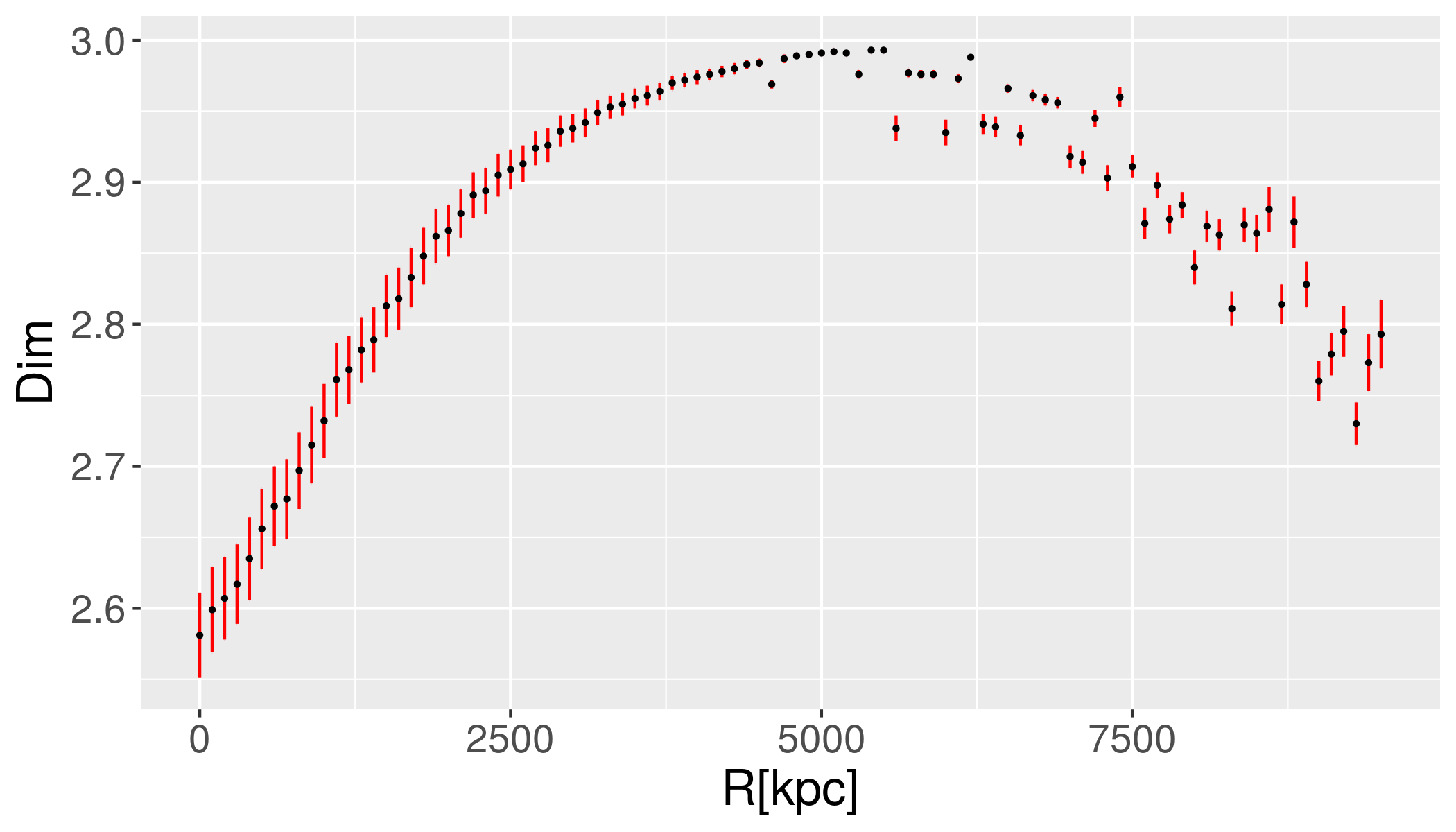}
	\caption{{Fractal dimension Dim as a function of position $R$}. }
 \label{dimension}
\end{figure}

For the milky way we first tested our simplest model of a spherical bulge of uniform density with fractional dimension $D=2.7$ and $D=2.3$, {which} correspond to the two limits found for the dimension (see Fig.~\ref{dimension}). We assumed a total mass of $M_{b+d}=6.29 \times 10^{10}~\text{M}_\odot$ and we adjusted the bulge radii $R_{b}$ to the data {of \citet{2016MNRAS.463.2623H,2017SoftX...6...54P}} using a nonlinear model on  Mathematica\textsuperscript{\textregistered}. We obtained $R_{b}=1.08~\text{kpc}$ with a {coefficient of determination $R^2\approx 0.943$, 0.926, and 0.898, for $D=2.3$, 2.7 and 3, respectively.} (see Fig.~\ref{milky}). 

\begin{figure}
	\centering
	\includegraphics[width=.42\textwidth]{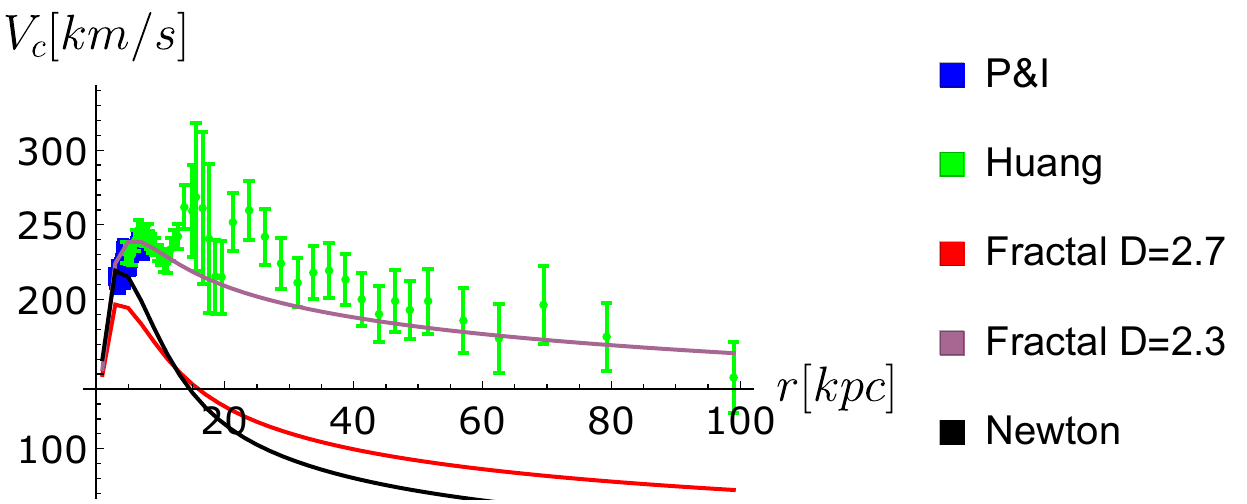}
	\caption{{Same as Fig.~\ref{milky}, but for the bulge-disk model.The mass of the bulge set to}
	$M_b=1.02 \times 10^{10}~\text{M}_\odot$ {and the mass of the disk to $M_d=5.27 \times 10^{10}~\text{M}_\odot$.}}
	\label{milkylocalfractal}
\end{figure}

To improve the fit we considered a bulge and disk model proposed {by \citet{2015Ap&SS.357...44S}}, adapted to our fractal form of the potential with a fractional dimension $D=2.3$ and $D=2.7$ (Sec.~\ref{dimensionmilky}). We considered the bulge and disk mass of $M_{b}=1.02 \times 10^{10}~\text{M}_\odot$ and $M_{d}=5.27 \times 10^{10}~\text{M}_\odot$ respectively {from \citet{Licquia_2015}} (see Fig.~\ref{milkylocalfractal}). {For the bulge-disk mdodel, we find $R^2\approx 0.984$, 0.942, and 0.916, for $D=2.3$, 2.7 and 3, respectively.}

It is important to {highlight} that in none of the models we considered the presence dark matter. 

\section{Conclusions}

{The uniform density model (Fig.~\ref{milky}) shows} a good fit for large radii for the case of $D=2.3${, but in no case it provides a good fit closer} to the center of the {Milky Way. {This model} is too simple to accurately} represent the {distribution} of matter in {our galaxy}. The mass we use for the milky way inside the sphere of constant density is the sum of the mass of the bulge and the disk  $M_{b+d}=6.29 \times 10^{10}~\text{M}_\odot$, with this mass we find a radii of $R_{b}=1.08~\text{kpc}$. The mass that we use is in the order of the expected mass for the total stellar mass \cite{Licquia_2015}. {On the other hand, the bulge and disk model (Fig.~\ref{milkylocalfractal}) shows} a better fit for $D=2.3${, compared to the simpler uniform model}. {In this case} we find a remarkably good fit not only for large radii but also for small ones. 
{As described in Sec.~\ref{milkyway_fit}, the overal goodness of fit of the different models was quantified computing their coefficient of determination $R^2$. The best performance measure is obtained by the bulge-disk $D=2.3$ model.}

%%%%%%%%%%%%%%%%%%%%%%%%%%%%%%%%%%%%%%%%%%%%%%%%%%%%%%%%%%%%%%%%%%%%%%%%%%%%%%
%                                                                            %
% Para figuras de dos columnas use \begin{figure*} ... \end{figure*}         %
%                                                                            %

\begin{acknowledgement}
The author thank N. Grandi and S. Hurtado. 
\end{acknowledgement}

%%%%%%%%%%%%%%%%%%%%%%%%%%%%%%%%%%%%%%%%%%%%%%%%%%%%%%%%%%%%%%%%%%%%%%%%%%%%%%
%                                                                            %
%  Por favor no modifique las líneas de la bibliografía, salvo el nombre     %
%  el archivo de Bibtex con la lista de citas (sin la extensión .BIB)        %
%                                                                            %
%  Please do not modify the following lines, except the name of the Bibtex   %
%  file (whithout the .BIB extension)                                        %
%                                                                            %
%%%%%%%%%%%%%%%%%%%%%%%%%%%%%%%%%%%%%%%%%%%%%%%%%%%%%%%%%%%%%%%%%%%%%%%%%%%%%% 

\bibliographystyle{baaa}
\small
\bibliography{bibliografia}

\begin{thebibliography}{14}
\providecommand{\natexlab}[1]{#1}

\bibitem[{{Bailer-Jones} et~al.(2018)}]{2018AJ....156...58B}
{Bailer-Jones} C.A.L., et~al., 2018, \aj, 156, 58

\bibitem[{{Elmegreen}(2000)}]{2000ApJ...530..277E}
{Elmegreen} B.G., 2000, \apj, 530, 277

\bibitem[{{Elmegreen} \& {Elmegreen}(2001)}]{2001AJ....121.1507E}
{Elmegreen} B.G., {Elmegreen} D.M., 2001, \aj, 121, 1507

\bibitem[{Gillessen et~al.(2009)}]{Gillessen_2009}
Gillessen S., et~al., 2009, ApJ, 692, 1075–1109

\bibitem[{{Huang} et~al.(2016)}]{2016MNRAS.463.2623H}
{Huang} Y., et~al., 2016, \mnras, 463, 2623

\bibitem[{Licquia \& Newman(2015)}]{Licquia_2015}
Licquia T.C., Newman J.A., 2015, ApJ, 806, 96

\bibitem[{Muslih \& Agrawal(2010)}]{Muslih}
Muslih S., Agrawal O., 2010, Int. J. Theor. Phys., 49, 270

\bibitem[{{Pato} \& {Iocco}(2017)}]{2017SoftX...6...54P}
{Pato} M., {Iocco} F., 2017, SoftwareX, 6, 54

\bibitem[{{Randall} \& {Sundrum}(1999)}]{randallsundrum}
{Randall} L., {Sundrum} R., 1999, \prl, 83, 3370

\bibitem[{Reid \& Brunthaler(2004)}]{Reid_2004}
Reid M.J., Brunthaler A., 2004, ApJ, 616, 872–884

\bibitem[{{Scelza} \& {Stabile}(2015)}]{2015Ap&SS.357...44S}
{Scelza} G., {Stabile} A., 2015, \apss, 357, 44

\bibitem[{Tarasov(2011)}]{TarasovBook}
Tarasov V., 2011, \textit{Fractional Dynamics}, Springer, Berlin.

\bibitem[{Tarasov(2005a)}]{TARASOV2005a}
Tarasov V.E., 2005a, Physics Letters A, 336, 167

\bibitem[{{Tarasov}(2005b)}]{TARASOV2005b}
{Tarasov} V.E., 2005b, Annals of Physics, 318, 286

\end{thebibliography}
 
\end{document}